\documentclass[%
mathleft,%
final,%
]{an}
\usepackage{graphicx}
\usepackage{times}
\usepackage{natbib}
\usepackage{balance}
\usepackage{url}
\bibpunct{(}{)}{;}{a}{}{,}
\def\apj{ApJ}%
\def\apjl{ApJ}%
\def\aap{A\&A}%
\def\mnras{MNRAS}%
\def\prd{Phys.~Rev.~D}%
\def\nat{Nature}%
\def\aplett{Astrophys.~Lett.}%
\def\memsai{Mem.~Soc.~Astron.~Italiana}%
\def\dmm{$\Delta \mu/\mu$}

\newcommand{\kms}{km\,s$^{-1}$}
\newcommand{\ms}{m\,s$^{-1}$}

\newcommand{\zabs}{\ensuremath{z_{\rm abs}}}

\newcommand{\daa}{\ensuremath{\Delta\alpha/\alpha}}

\newcommand{\systemARedshift}{1.6919}

\newcommand{\ReportStatisticalError}[2]{\ensuremath{#1\,\pm\,#2_{\rm stat}}}

\newcommand{\psystemAdaaOnePlace}{-3.8}

\newcommand{\psystemAdaaStatisticalErrorOnePlace}{2.1}

\newcommand{\psystemAsidamdaaTwoPlaces}{+1.14}

\newcommand{\psystemAsidamdaaStatisticalErrorTwoPlaces}{2.58}

\begin{document}

\Pagespan{1}{}
\Yearpublication{2014}%
\Yearsubmission{2013}%
\Month{1}%
\Volume{335}%
\Issue{1}%
\DOI{This.is/not.aDOI}%

\title{Fundamental constants  and high resolution spectroscopy\thanks{based on 
observations obtained with UVES at the the 8.2m Kueyen ESO
telescope programme L185.A-0745}}
\author{P. Bonifacio\inst{1}\fnmsep\thanks{Corresponding author:
  \email{Piercarlo.Bonifacio@obspm.fr}\newline}
\and H. Rahmani\inst{2}
\and J. B. Whitmore\inst{3} 
\and M. Wendt \inst{4,5}
\and M. Centurion\inst{6}
\and  P. Molaro\inst{6}
\and R. Srianand\inst{2}
\and M. T. Murphy\inst{3}
\and P. Petitjean\inst{7}
\and I. I. Agafonova \inst{8}
\and S. D'Odorico\inst{9}
\and T. M. Evans \inst{3}
\and S. A. Levshakov\inst{8,10}
\and S. Lopez\inst{11} 
\and C. J. A. P. Martins\inst{12}
\and D. Reimers\inst{4}
\and G. Vladilo\inst{6}
 }
\titlerunning{Variations of $\alpha$ from high resolution spectroscopy}
\authorrunning{P. Bonifacio et al.}
\institute{
GEPI, Observatoire de Paris, CNRS, Univ. Paris Diderot, Place Jules Janssen, 92190 Meudon, France
\and Inter-University Centre for Astronomy and Astrophysics, Post Bag 4, Ganeshkhind, 411 007 Pune, India 
\and Centre for Astrophysics and Supercomputing, Swinburne University of Technology, Hawthorn, VIC 3122, Australia
\and Hamburger Sternwarte, Universit\"at Hamburg, Gojenbergsweg 112, 21029 Hamburg, Germany
\and Institut f\"ur Physik und Astronomie, Universit\"at Potsdam, 14476 Golm, Germany
\and Istituto Nazionale di Astrofisica,
Osservatorio Astronomico di Trieste,  Via Tiepolo 11,
34143 Trieste, Italy
\and Universite Paris 6, Institut d'Astrophysique de Paris, CNRS UMR 7095, 98bis bd Arago, 75014 Paris, France
\and Ioffe Physical-Technical Institute, Polytekhnicheskaya, Str. 26, 194021 Saint Petersburg, Russia
\and ESO Karl, Schwarzschild-Str. 2 85741 Garching, Germany
\and St.~Petersburg Electrotechnical University `LETI', Prof. Popov Str. 5,
197376 St.~Petersburg, Russia
\and Departamento de Astronomia, Universidad de Chile, Casilla 36-D, Santiago, Chile
\and Centro de Astrof\'isica, Universidade do Porto, Rua das Estrelas, 4150-762 Porto, Portugal
}
\received{30 Aug 2013}
\accepted{   Jan 2014}
\publonline{later}
\abstract{Absorption-line systems detected in high resolution quasar spectra can be used to 
compare the value of dimensionless fundamental constants such as the fine-structure constant,\ 
$\alpha$ , and  
the  proton-to-electron mass ratio, $\mu  = m_p/m_e$, 
as measured in remote regions of the Universe
to their value today on Earth. 
In recent years, some evidence has emerged of small temporal and also spatial 
variations in $\alpha$ on cosmological scales which  may reach a fractional level of 
$\approx$  10 ppm (parts per million). 
We are conducting a Large Programme of observations with the Very Large Telescope's Ultraviolet and Visual Echelle Spectrograph (UVES), and are obtaining high-resolution (R $\approx$ 60 000) and high signal-to-noise ratio (S/N $\approx$ 100) spectra calibrated specifically 
to study the variations of the fundamental constants. 
We here  provide a general overview of the Large Programme and
report   on the first results  for   
these two constants, discussed in detail in Molaro et al. and Rahmani et al.
A stringent bound   for $\daa$  is obtained for the absorber at 
$\mathrm z_{abs} = 1.6919$ towards HE 2217-2818. 
The absorption profile is complex with several very narrow features, and is modeled  with  
32 velocity components.  The relative variation in $\alpha$ in this system is  
 $+1.3\pm   2.4_{\mathrm stat} \pm 1.0_{\mathrm sys}$ ppm if \ion{Al}{ii} $\lambda$ 1670\,\AA\
and three \ion{Fe}{ii} transitions  are used,  
and   +1.1 $\pm  2.6_{\mathrm stat}\, $ppm in a 
slightly different analysis with  only  \ion{Fe}{ii} transitions  used.   
This is one of the tightest  bounds on $\alpha$-variation from 
an individual absorber and  reveals no evidence for
variation in $\alpha$ at the 3-ppm precision level (1-$\sigma$
confidence).   The expectation at this sky
position of the recently-reported  dipolar variation of
$\alpha$ is $(3.2$--$5.4)\pm1.7$\,ppm depending on dipole model
used and  this  constraint of
$\daa$   at face value is not supporting   this expectation but not 
inconsistent with it at the 3$\sigma$ level. 
For the proton-to-electron mass ratio the   analysis of the $H_2$  
absorption lines of  the $z_{abs}  \approx$ 2.4018 damped Ly$\alpha$ system towards 
HE 0027- 1836  provides  $ {\Delta}{\mu}/{\mu} = (-7.6 \pm  8.1_{\mathrm stat} \pm  6.3_{\mathrm sys}) $ ppm 
which is also consistent with a null variation. The cross-correlation analysis between  individual 
exposures taken over three years  and comparison with almost simultaneous  
asteroid observations  revealed  the presence of a possible wavelength dependent velocity drift  
as well as of inter-order distortions which probably 
dominate the systematic error and are a significant obstacle to achieve more accurate measurements.
 }
\maketitle
\sloppy
\section{Introduction}
General relativity and the standard model of particle physics depend on a number of  independent numerical parameters that determine the strengths of the different forces and the relative masses of all known fundamental particles. There is no theoretical explanation of why they have the  values  they have but they  determine the properties of atoms, cells, stars and the whole Universe. They are commonly referred to as the fundamental constants of Nature,  but  a variation of the constants, at some level, is a common prediction of most modern extensions of  the Standard Model (see Uzan 2003 for a review). 
That physical constants could vary over cosmological
time is an idea that has been around ever since
Dirac's ``Large Number Hypothesis'' \citep{dirac}.
It is currently of great interest in the context of
cosmologically relevant scalar fields,  like 
quintessence (see
\citealt{quint} for a review).
An attractive implication of quintessence models for the dark energy is that the rolling scalar field  producing a negative pressure and therefore the  acceleration of the universe may couple with other fields and be revealed by a change in the fundamental constants \citep{amendola}.
Variation of the fundamental constants is foreseen also in other theories beyond the standard model. For instance,    in  theories  involving more than four space-time dimensions the constants we observe are merely four-dimensional shadows of the truly fundamental high dimensional constants and they may  be seen to vary as the extra dimensions change slowly in size during their cosmological evolution.   
 The fine structure constant $\alpha = e^2/\hbar c$
is dimensionless and governs the coupling between 
photons and electrons. By solving the Schr\"odinger equation
for the hydrogen atom, the bound states are given by
\begin{equation}
E_{n} = -\alpha ^2 {mc^2\over 2 n^2}
\end{equation}

\noindent where $n$ is the principal quantum number (n=1,2,...,$\infty$)
and $\alpha$ is the above defined fine structure constant
\citep[][p. 354 eq. 17]{messiah1}.
When the relativistic
corrections are considered  the 
eigenvalues corresponding to angular momentum $J$
and principal quantum number $n$ can be approximated to

\begin{equation}
E_{nJ} = mc^2\left [ 1 + {\alpha^2\over (n-\epsilon_J)^2} \right ] ^{-1/2}
\end{equation}

\noindent 
where $\epsilon_J$ is a function of $J$ and $\alpha^2$ \citep[][p. 802, eq. 179]{messiah2}. 
Whenever we have a fine-structure multiplet, i.e.
transitions between energy levels with the same
principal quantum number and different $J$, the relativistic
corrections are  proportional to $\alpha^2$, to first order, 
as can be seen by doing a power series expansion of the term in square brackets
in eq. 2. 

The simplest case  is that of alkali 
doublets such as  \ion{Li}{i}, \ion{Na}{i}, \ion{K}{i},
but also of the alkali ions \ion{C}{iv} and \ion{Si}{iv} where 
the splitting of the doublet, i.e. the wavelength
separation of the two components is a function of
$\alpha$.  By 
measuring  the alkali  splitting in gas at redshift $z$ we can
measure the value of $\alpha$ at a different instant
of space-time. This means we can effectively  probe the variations
of $\alpha$ over space-time. 
 Earth-based laboratories   have so far revealed no variation in their values. For example, the constancy of the fine structure constant  stability is ensured to within a few parts per 10$^{-17}$  over a 1 yr period  (Rosenband et al. 2008). Hence its status as  truly  constants  is amply justified.  Astronomy has a great potential in probing their variability  at very large distances  and in the early Universe. 

The first attempts to measure variation of $\alpha$
using QSO spectra \citep{savedoff,bahcall} could only
achieve an accuracy of $10^{-2}$ in $\Delta\alpha/\alpha$.

However,  the transition frequencies of the narrow metal absorption lines observed in the spectra of distant  quasars    are  sensitive to $\alpha$. 
Thus
the many-multiplet (MM)
method has been introduced, which allows all observed transitions to
be compared, gaining access to the typically much larger dependence of
the ground state energy levels on $\alpha$ \citep{Dzuba:1999:888}.  Overall,
the MM method improves the sensitivity to the measurement of a
variation of $\alpha$ by more than an order of magnitude over the
alkali-doublet method. 

The change in the rest-frame frequencies between the laboratory,
$\omega_{i}(0)$, and in an absorber at redshift $z$, $\omega_{i}(z)$,
due to a small variation in $\alpha$, i.e.~\daa $\ll 1$, is
proportional to a $q$-coefficient for that transition:

\begin{equation}\label{eq:da1}
\omega_{i}(z) \equiv \omega_{i}(0) + q_i\left[\left(\alpha_z/\alpha_0\right)^2-1\right]\,,
\end{equation}

\noindent 
where $\alpha_0$ and $\alpha_z$ are the laboratory and absorber values
of $\alpha$, respectively \citep{Dzuba:1999:888}.  
  The change in frequency is observable as a
velocity shift, $\Delta v_i$, of the $i-th$ transition.

\begin{equation}\label{eq:da2}
\hspace{1em}\frac{\Delta v_i}{c} \approx -2\frac{\Delta\alpha}{\alpha}\frac{q_i}{\omega_{i}(0)}\,.
\end{equation}

The MM method is based on the comparison of measured velocity shifts
from several transitions having different $q$-coefficients to compute
the best-fitting \daa.

The MM method  and the
advent of 8m class telescopes  that could provide
high resolution spectra of QSOs  gave the first hints that the fine 
structure constant might change its value over time, being lower in the past 
by about 6 part per million (ppm)  (Webb et al. 1999, Murphy et al. 2004). 
With the addition of other 143  VLT-UVES absorbers  
Webb and collaborators  arrived at the surprising
conclusion that although on average
there is no variation of $\alpha$ there are significant
variations along certain directions in the sky. 
They have  found  a   4-$\sigma$ evidence for a dipole-like 
variation in $\alpha$  across the sky at the 10 ppm level 
(Webb et al. 2011; King et al. 2012).  Several other constraints
from higher-quality spectra of individual absorbers exist
\citep{chand2006,lev2007}
 but none directly 
support or strongly conflict with the $\alpha$ dipole 
evidence and a possible systematic producing opposite values 
in the two hemispheres is not easy to identify.

The 
proton-to-electron mass ratio, $\mu$, is also a   dimensionless constant
which can be probed  experimentally. 
It is known that the wavelengths of the rovibronic  
molecular transitions are sensitive to $\mu$. 
In a diatomic molecule the energy of the rotational transitions 
is proportional to the reduced mass 
of the molecule, $M$, and that of vibrational transitions is proportional  
to $\sqrt{M}$, in the first order approximation. The frequency of 
the rovibronic transitions in Born-Oppenheimer approximation 
can be written as,
\begin{equation}
\nu = c_{\rm elec} + c_{\rm vib} / \sqrt{\mu} + c_{\rm rot} / \mu
\end{equation}
where $c_{\rm elec}$, $c_{\rm vib}$, and $c_{\rm rot}$ are some numerical coefficients 
related, respectively, to electronic, vibrational and  rotational transitions. 
Therefore, by comparing the wavelength of the molecular transitions detected in 
quasar spectra with their laboratory values one can measure the 
variation in $\mu$ (i.e. \dmm\ $\equiv (\mu_z-\mu_0)/\mu_0$ where $\mu_z$ and $\mu_0$ are 
the values of proton-to-electron mass ratio at redshift $z$ and today) over cosmological time scales. 
Using  intervening molecular absorption lines seen in the high-$z$
quasar spectra for measuring  \dmm\ in the distant universe was first proposed 
by \citet{Thompson75}. As H$_2$ is the most abundant 
molecule its Lyman and Werner absorption lines seen in the quasar absorption 
spectra have been frequently used to constrain the variation of $\mu$.
However, H$_2$ molecules are 
detected  in only a few percent of
the high redshift damped Lyman-$\alpha$ (DLA) systems 
\citep{Srianand12}
with only a handful 
of them being suitable for probing the variation of $\mu$.

If $\mu$ varies, the observed wavelengths of different H$_2$ lines will shift
differently with respect to their expected wavelengths based on laboratory
measurements and the absorption redshift. The sensitivity of the wavelength of 
the i'th H$_2$ transition to the variation 
of $\mu$ is generally parametrised as
\begin{equation}
\lambda_i = \lambda_i^0 (1+z_{\rm abs})\big{(}1+K_i\frac{\Delta\mu}{\mu}\big{)},
\label{eq_dm}
\end{equation}
where $\lambda_i^0$ is the rest frame wavelength of the transition, $\lambda_i$  
is the observed wavelength, $K_i$ is the sensitivity coefficient 
of i'th transition, and \zabs\ is the redshift of the H$_2$ absorber. 
Alternatively  Eq. \ref{eq_dm} can be written as 
\begin{equation}
z_i = z_{\rm abs} + C K_i, ~~~~~ C = (1+z_{\rm abs})\frac{\Delta\mu}{\mu}
\label{eq_dm_a}
\end{equation}
which clearly shows that \zabs\ is only the mean redshift of transitions with $K_i$ = 0. 
Eq. \ref{eq_dm_a}  is sometimes presented as 
\begin{equation}
z_{\rm red} \equiv \frac{(z_i - z_{\rm abs})}{(1 + z_{\rm abs})} = K_i \frac{\Delta\mu}{\mu}
\label{eq_dm_b}
\end{equation}
that shows the value of \dmm\ can be determined using a linear regression analysis of reduced 
redshift ($z_{\rm red}$) vs $K_i$. 
This method has been frequently  used in the literature for constraining the variation 
of $\mu$ \citep[see][]{Varshalovich93,Cowie95,Levshakov02mnras,malec,wm11,vw11,wm12}.
However, at present  measurements of \dmm\ using $H_2$ is limited 
to 6 $H_2$-bearing DLAs at $z \ge$ 2. All of these analyses suggest that
$|$\dmm$|\le10^{-5}$ at $2\le z \le 3$. The best reported constraints
based on a single system being \dmm\ = $+(0.3\pm3.7)\times 10^{-6}$
reported by \citet{King11} towards Q~0528$-$250.  Among the developments in the field since then 
we must signal
a number of high precision measurements 
of $\Delta\mu/\mu$ both using UV lines
\citep{malec,wm11,vw11,wm12}

At $z\le 1.0$ a stringent constraint on \dmm\ is obtained using inversion transitions
of NH$_3$ and rotational molecular transitions \citep{Murphy08,Henkel09,Kanekar11}. 
The best reported limit using this technique is \dmm\ = $-(3.5\pm1.2)\times 10^{-7}$ \citep{Kanekar11}. 
\citet{Bagdonaite13} obtained the strongest
constraint to date of \dmm\ = $(0.0\pm1.0)\times10^{-7}$ at $z = 0.89$
using methanol transitions.
In the Galaxy  stringent bounds have been obtained in the millimetre and sub-millimetre   
domain by \citep{l10a,l10b,l10c}.

However, \dmm\ measurements using NH$_3$ and
CH$_3$OH are restricted to only two specific systems at $z\le 1$. Alternatively
one can place good constraints using 21-cm absorption in conjunction
with metal lines and assuming all other constants have not changed. \citet{Rahmani12} have
obtained \dmm = $(0.0\pm1.50)\times 10^{-6}$ using a well selected sample
of four 21-cm absorbers at \zabs $\sim$1.3. \citet{Srianand10} have 
obtained \dmm = $(-1.7\pm1.7)\times10^{-6}$ at $z \sim$3.17 using 
the 21-cm absorber towards J1337$+$3152. 
However, one of the main systematic uncertainties in this method comes from 
how one associates 21-cm and optical absorption components.

 An ESO Large Programme (LP)
has been undertaken in the last four semesters to address the case of constant variability.
 We shall here briefly  describe the programme, 
and show   some of its first
results.

\begin{table}
\caption{QSO targets of the Large Programme.\label{targets}}
\centering
\begin{tabular}{lllll}
\hline
QSO & $\alpha$(2000) &  $\delta$(2000) & $V$ &\\
\hline
HE 0002--4214    & 00 04 48.20&--41 57 28.0&17.2\\
HE 0027--1836    & 00 30 23.63&--18 19 56.0&17.9\\ 
QSO J0120+2133  & 01 20 17.26& +21 33 46.4&16.1\\
PKS 0237--23     & 02 40 08.17&--23 09 15.75&16.6\\
QSO J0407--4410  & 04 07 17.99&--44 10 13.4&17.6\\
QSO J0455--4216  & 04 55 22.90&--42 16 16.9&17.1\\
HE  0940--1050   & 09 42 53.49&--11 04 25.9&16.6\\
QSO J1215--0034  & 12 15 49.81&--00 34 32.2&17.5\\
QSO J1333+1649  & 13 33 35.78& +16 49 04.0&16.7\\
HE 1341--1020    & 13 44 27.10&--10 35 42.0&17.1\\
HE 1347--2457    & 13 50 38.88&--25 12 16.7&16.3\\
QSO J1549+1911  & 15 51 52.48& +19 11 04.2&15.8\\
QSO J2136--4308  & 21 36 06.04&--43 08 18.1&17.7\\
HE 2217--2818   & 22 20 06.77&--28 03 23.4&16.0\\
QSO J2208--1944  & 22 08 52.07&--19 44 00.0&17.3\\
\hline
\end{tabular}
\end{table}

\section{The UVES Large Programme}

The main drawback of the sample assembled by 
\citet{webb} is that the observations where
mainly acquired  with scientific objectives other
than the measurement of $\Delta\alpha/\alpha$
thus systematic errors are not 
monitored and minimised. 
In 2010  our  Large Program
of optical observations dedicated to 
measuring $\alpha$ and $\mu$ in distant galaxies was approved
by the ESO Observing Programmes Committee.   
The  Large Program was
granted  42  nights beginning mid 2010 at UVES at the ESO VLT to obtain  
a high-quality sample of quasar spectra, calibrated specifically for the purpose
of constraining variations  in $\alpha$ and $\mu$   
to the ultimate precision allowed by current technology.
For the first time the spectra were observed primarily
for this purpose, with the explicit aim to keep
calibration errors under control. 
The fundamental physical questions being addressed demand a level of rigour in quasar absorption
studies well beyond the norm 
previously adopted.  

The signal-to-noise ratio of quasar spectra 
is one of the  main factors  in the error budget. This, in turn, 
limits one's ability to track systematic errors.
However, by careful   selection of targets our Large Program focuses on

\begin{itemize}
\item 15  among the brightest known quasars showing a suitable absorber
\item a relatively large number of absorbers  along their sight-lines: 22 in total.
\end{itemize}

The coordinates and magnitude of the target QSOs are given in Table \ref{targets}.
 This means we have observed
each absorber for more than 
three nights on average, which allowed us  to build for many absorbers a much higher 
signal-to-noise ratio 
than achieved in all previous studies except the 
two ``test case'' absorbers studied in \cite{Molaro:2008:173}. In these
cases the photon statistical noise was reduced well below that from systematic errors. 
Our Large Programme 
achieved this for all relevant absorbers.
For each absorber we have a high  enough signal-to-noise ratio 
to convincingly detect, model and remove any remaining systematic 
errors down to the level of few  ppm,  
thereby allowing a convincing detection of any variation 
in $\alpha$ at the level seen
in the Keck spectra \citep{Murphy:2003:609}.

The measurements rely on detecting a pattern of small relative wavelength shifts between different
transitions spread throughout the spectrum. Normally, quasar spectra are calibrated by comparison with spectra
of a hollow cathode lamp (normally thorium) rich in unresolved spectral lines. 
However since the lamp is located inside the
spectrograph, the calibration light 
traverses a slightly different optical path with respect
to the quasar light, so the comparison
is not perfect . The Large Program adopts several 
innovations to ensure that we achieve the ultimate precision available:

\begin{itemize}
\item  we  systematically observed bright asteroids, whose reflected sunlight 
spectra contain many spectral features, 
typically narrower and sharper than QSO absorption lines. 
These observations allow to generate a transfer function  for correcting the comparison lamp
wavelength scale. This technique was recently pioneered by  us \citep{Molaro:2008:559}

\item  we  observed bright stars through an iodine gas 
absorption cell, as done for extrasolar planet searches, providing an even more precise
transfer function for part of the wavelength range,  important for varying constants; 

\item  we took a series of lamp exposures bracketing 
the quasar exposures to ensure the best possible starting
point for this transfer function.
\end{itemize}

Previous estimates of wavelength calibration errors in varying measurements are at 
the 3 - 5 ppm level \cite{Molaro:2008:173}. With the three innovative approaches above, we
expect to suppress/remove them below the 1 ppm level for individual quasar absorbers.

\subsection{Systematic effects in the wavelength calibration}

A major step forward towards the understanding of the systematic
effects that limit the precision of wavelength calibration
has been achieved by the use of a Laser Frequency
Comb (LFC) on the HARPS spectrograph \citep{Wilken:2010:L16,Wilken}.
These observations were capable of highlighting the presence
of tiny differences in the pixel sizes of the CCD detectors,
that are due to the manufacturing process. Quite interestingly the
list of wavelengths of the Th-Ar lamp normally used to
calibrate HARPS  \citep{Lovis:2007:1115}, has the inaccuracies 
of $\pm$40\,\ms\ due to the detector, folded in, thus when
this line list is used as reference one should expect 
locally errors of this order of magnitude. 
No experiment with an LFC has been carried out so far on the
spectrographs on 8m class telescopes, such as UVES or HIRES, 
yet pixel size differences of the same order as those
found in HARPS should be expected for these detectors too.
\citet{Griest:2010:158} and \citet{Whitmore:2010:89} 
compared the calibration obtained with the Th-Ar lamp with
that obtained from an absorption cell of molecular iodine,
for HIRES and UVES, respectively. In both cases they were able
to highlight distortions of the wavelength scale with a 
jig-saw pattern and peak-to-peak amplitude of several hundreds \ms\
along the echelle orders.
Although the origin of these distortion is not completely
elucidated  it is likely due to  a combination of inhomogeneity
in pixel size of the detectors and errors in the reference 
line positions of the Th-Ar lamps.

\subsection{Solar-asteroid comparison}\label{sol-ast}

Comparison of different calibration laboratory sources, like Th-Ar, LFC and 
$\mathrm I_2$ cell  helps to better characterize 
the systematics of our wavelength scale. A complementary and
very interesting technique is the use of the spectrum of
an astronomical object. 
This has the advantage that the light follows exactly the
same optical path as  the scientific target.
A very attractive astronomical source for wavelength
calibration are asteroids \citep{Zwitter}, that reflect
the solar spectrum, imprinting on it minor signatures, 
mainly broad and shallow abosptions, and that have
radial velocities that are known, from their orbital
solution, to an accuracy of a few \ms .
To test the accuracy of our wavelength scale
one possibility is to compare the measured 
line positions in an asteroid spectrum
with those from a solar atlas obtained
with a different instrument.  
A frequently used solar atlas  for this
purpose is the Kurucz solar flux spectrum
\citep{Kurucz05}\footnote{
\url{http://kurucz.harvard.edu/sun/fluxatlas2005}}. 
The 
wavelength scale
of this atlas is corrected for the gravitational redshift ($\sim$ 0.63 \kms).
The claimed accuracy
of the absolute wavelength scale 
 is $\sim 100$ ms$^{-1}$ \citep{Kurucz05}, However, this
should be taken as an average value, since comparison 
of individual lines with synthetic spectra computed from
hydrodynamical models of the stellar photosphere, that take
into account the convective shifts, may show deviations
as large as several hundreds \ms\ \citep[see e.g.][]{caffau}.

\begin{figure} 
\centering
\resizebox{\hsize}{!}{\includegraphics[angle=90]{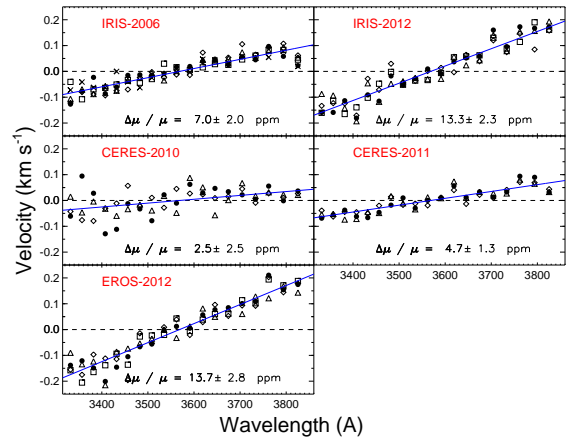}}
\caption{The velocity shift measurements using cross-correlation analysis between solar and 
asteroids spectra. The solid line in each panel shows the fitted line 
to the velocities. The \dmm\ corresponding to the slope of the fitted straight line is 
also given in each panel. 
}
\label{fig_sol_ast}
\end{figure}

Following this approach  \citet{Molaro08} 
compared the positions of individual lines measured
in the spectra of asteroids and in the solar atlas. 
This is not possible in the near ultra-violet, 
where the line blending in  the solar spectrum
is so high that positions of individual lines
cannot be measured.
An alternative way to perform this comparison
has been explored by our group in  \citet{Rahmani2013}
where we cross-correlated the spectra of asteroids, corrected
for their radial velocity,
observed with UVES over several years with the
solar atlas. This technique allows to highlight
the presence of wavelength-dependent  velocity offsets 
between the asteroid spectrum and the solar atlas.
We show in  
Fig.~\ref{fig_sol_ast} 
the results of the analysis of \citet{Rahmani2013}, where 
different spectra of the same asteroid observed at 
a different epochs are shown as 
different symbols.  
It is clear from the figure that the offsets increase
with wavelength, but the slope is not the same at all
epochs, being larger for the asteroids observed in 
2012.

From our point of view it is important to assess
the effect of these offsets on a measurement
of the variation of a constant, such as \daa\ or
\dmm , assuming that the offsets seen in the
asteroid spectra are the same in the QSO spectra.

In Fig. \ref{fig_sol_ast}, taken from \citet{Rahmani2013},
an intrinsic \dmm = 0 is assumed and the H$_2$ lines are
assumed to be imprinted on the spectrum, displaying the
measured velocity offset.  
The estimated \dmm\ , assuming  all the velocity 
differences to be due to a variation in $\mu$ is then 
given in each panel.  
It is striking that at least in two cases one
would conclude on the existence of a variation
in $\mu$ at a level of 4.5$\sigma$.  
It is thus crucial to detect and remove such offsets
before analysing the QSO data, to avoid spurious detections.

 \citet{Molaro:2011:167} and \citet{Whitmore2013} compared solar features observed both with
HARPS and UVES and found such `intra-order distortions' in the UVES
spectrum.   In HARPS the
offsets were measured up to 50 \ms within one order and in UVES, where
the pixel size is a factor of three larger, the offsets are found  a factor 
of three 
larger.  

 \section{\daa  ~ towards   HE\,2217$-$2818 }

\begin{figure*}
\begin{center}
\includegraphics[width=0.75\textwidth]{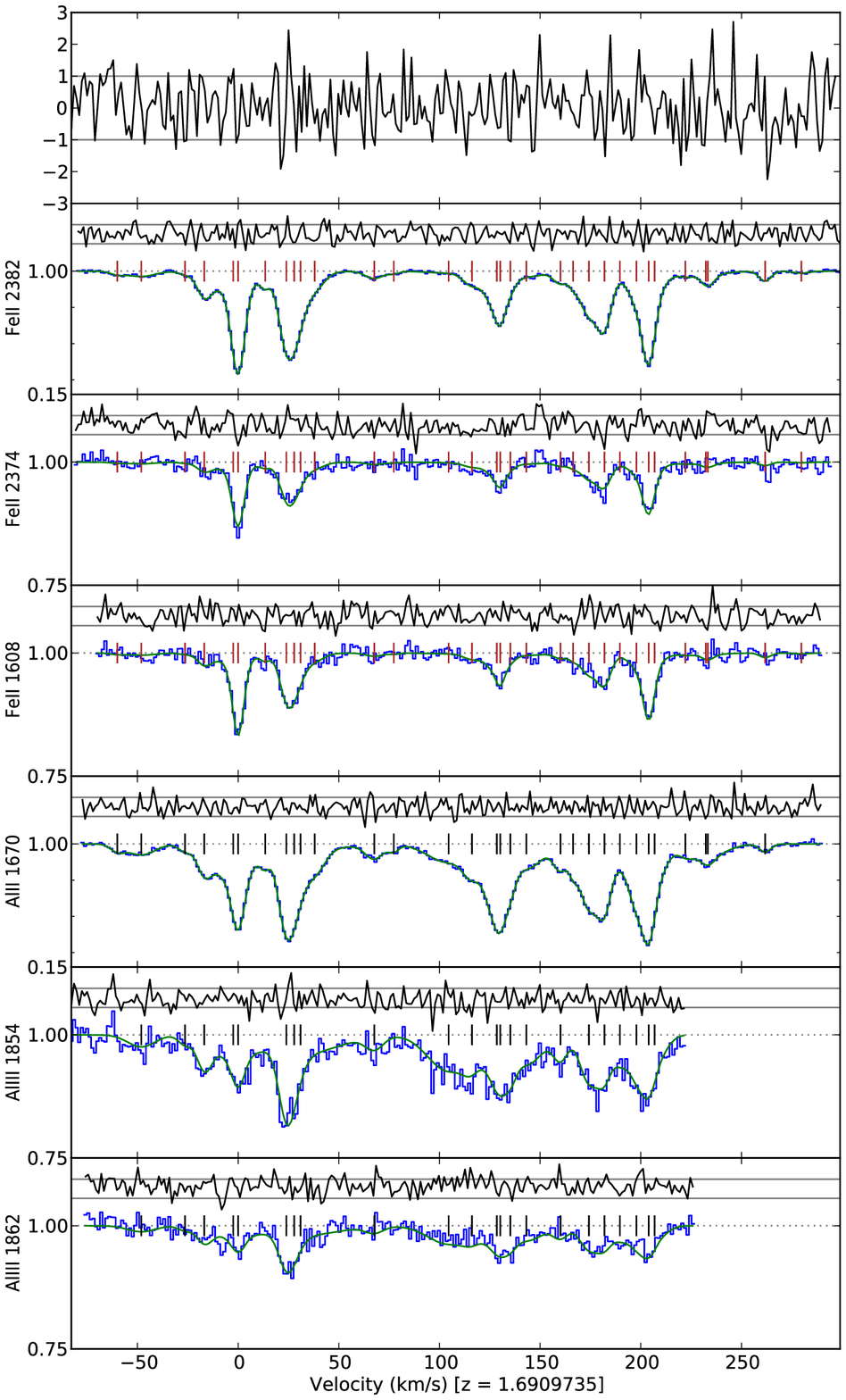}
\caption{Transitions in absorption system at $\zabs=\systemARedshift$ used to
  derive \daa\ in our second analysis approach. The Voigt profile
  model (green line) is plotted over the data (blue histogram). The
  velocity of each fitted component is marked with a vertical line and
  the residuals between the data and model, normalised by the error
  spectrum, are shown above each transition. The top panel shows the
  composite residual spectrum -- the mean spectrum of the normalised
  residuals for all transitions shown -- in units of $\sigma$. 
Credit: Molaro et al. A\&A 555, 68, 2013
reproduced  
with permission, \copyright ESO}
\label{fig:sys169}
\end{center}
\end{figure*}

The first result of our Large Programme
is the analysis of \daa\ in the absorption 
systems towards 
HE\,2217$-$2818 and has been presented
in \citet{mol13}. We refer the reader to that paper
for all the details of this analysis, 
that is  here summarized.
Of the five potentially useful absorption systems
only the one   at $z_{abs}=1.6919$ provides
a tight bound on \daa.
In spite of the fact that the system is complex,
constituted of several sub-components that span
about 250 \kms , each sub-component is narrow
enough to allow a precise determination of
its wavelength.
A matter of concern are the telluric absorptions, that
are imprinted on the spectrum and can seriously
affect the measured wavelength of the intergalactic
absorptions. The telluric lines were identified with the
help of the spectrum of a hot, fast rotating star.
No attempt was made to remove the telluric absorptions, 
two different approaches were adopted to deal with them.
In the first case any
intergalactic absorption affected by telluric lines 
was removed
from the analysis, in the second case
the portion of the spectra affected were masked and 
not considered in the analysis.
In Fig.\ref{fig:sys169}, reproduced from \citet{mol13},
the six ``clean'' lines that are used in the  first case 
are shown together with the best fitting (in the $\chi^2$ sense) 
model. 
The best fitting model shown includes as many as 32 sub-components
for each transition.
The number of components was determined by iteratively fitting the profiles with an increasing number of components, until a minimum in the reduced $\chi^2_{\nu}$ was obtained.

The best-fit provides  
 \daa=  $+1.3\pm   2.4_{ stat} \pm 1.0_{sys}$  {\rm ppm}   
     
In the second approach, in which a larger number of transitions
is considered, acceptable fits can be obtained with a slightly 
smaller number of components, thirty, rather than thirty-two.
It is nevertheless reassuring that the two
approaches yield consistent results, within 
our estimated statistical error, supporting
the robustness of our analysis: 
$  \daa=
    \ReportStatisticalError{\psystemAdaaOnePlace}
      {\psystemAdaaStatisticalErrorOnePlace}\,{\rm ppm} 
     $ for the second   approach.
     
     One matter of concern is the use of different ions, 
given that ionization effects may introduce a systematic effect
in the \daa\ measurements
\citep[e.g.][]{Levshakov:2005:827}. In this system   we can use as many as six \ion{Fe}{ii}
transitions, which have   different   q coefficients  making it feasible to perform an analysis of \daa\
based on this ion only.
Within the second approach this  leads to
 $\daa = 
  \ReportStatisticalError{\psystemAsidamdaaTwoPlaces}
  {\psystemAsidamdaaStatisticalErrorTwoPlaces}$\,ppm,
which is, again, statistically consistent with the  
other two analysis. 

\subsection{Implications for the spatial dipole in \daa}

Our results are consistent with no variation 
in $\alpha$ along the line of sight to 
HE\,2217$-$2818, the system at $z_{abs}= 1.6919$
with a very stringent bound, but the other 
five systems at lower redshifts are consistent with
this conclusion. 
It is interesting to compare this null result with
the prediction of the dipole model for the
spatial variation of \daa .
We consider the model proposed 
by \citet{King:2012:3370}, that stems
from the analysis of nearly 200 measurements
obtained both with UVES at VLT and HIRES at Keck.
The 
combined data, at  an approximate
mean  redshift $\ge$1.8, suggests a spatial  variation of \daa\
that can be described by the sum of
a monopole and a dipole 
in the direction with equatorial coordinates 
$\mathrm 17.3h\pm1.0h$, $-61^\circ\pm10^\circ$ 
\citep{King:2012:3370}. 
This can also be described by a simpler model, 
with only the dipole term 
in the direction 
$\mathrm 17.4h\pm0.9h$, $-58^\circ\pm9^\circ$  \citep{King:2012:3370}.
For the line of sight towards
HE\,2217$-$2818 
the simple dipole-only model 
predicts $\daa=+5.4\pm1.7$\,ppm.  
Thus our measurement 
differs from the simple dipole prediction 
by 1.3\,$\sigma$. The corresponding
prediction for the monopole plus dipole model
is $\daa=+3.2\pm1.7$\,ppm.
Our null result does not support the existence of
the dipole, yet it is not stringent enough
to rule it out.

\begin{figure} 
\centering
\resizebox{\hsize}{!}{\includegraphics[angle=90,clip=true]{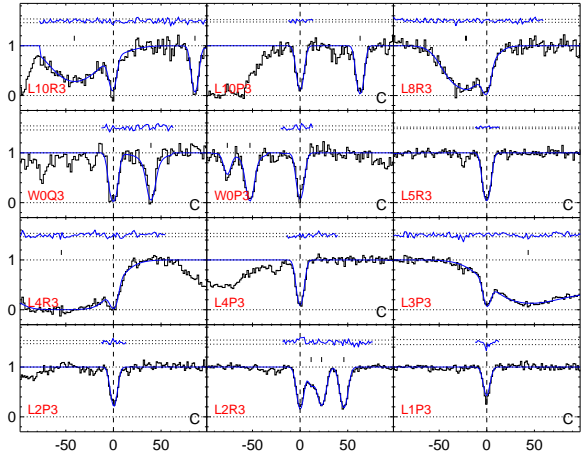}}
\caption{Absorption profile of $H_2$ transitions of $J$ = 3 level and the best 
fitted Voigt profile to the combined spectrum. The normalized residual 
(i.e.([data]$-$[model])/[error]) for each 
fit is also shown in the top of each panel along with the 1$\sigma$  line. 
We  identify the clean absorption lines by  using 
the letter "C" in the right bottom of these transitions.  The vertical ticks mark the 
positions of fitted contamination. Reproduced from \citet{Rahmani2013}, with permission.}
\label{fig_J3}
\end{figure}

\section{\dmm\ towards  HE 0027-1836}

 The  DLA at $z = $ 2.4018 towards HE 0027-1836 shows a  H$_2$ cloud  with    over 100 H$_{2}$ 
lines in the  observed wavelength range of 
3330 \AA\ to 3800 \AA\ which can effectively be used to probe $\mu$.
The detected lines are  from different 
rotational states  ( $0 \le J\le 6 $) and have 
  a wide range of oscillator strengths thus 
allowing a  very accurate  
modeling of the molecular  cloud. 
The analysis of this system has been 
reported by \citet{Rahmani2013} and we refer
the reader to that paper for all the details,
we here summarize the main results.

From all the detected 
 absorptions   71 strong and relatively unblended H$_2$ lines 
were selected for the analysis. 
To derive  \dmm\  
we used either: i)  a  linear regression analysis  between the line redshift $z_{\rm red}$ and its sensitivity coefficient  $K_i$  
\citep[cfr][]{wm10,wm11} or  ii) by detailed modeling of the lines   inserting   \dmm\ as an 
additional parameter  \citep[cfr][]{King11}.

For the former method we obtained the redshifts of individual transitions  from 
\textsc{vpfit}.  An example of the  fitted 
Voigt profile  for $J$ = 3 transitions is provided in  Figure \ref{fig_J3}. 
Since differences
in the excitation temperatures and  broadening  
between  high and low $J$-levels
may be due to  different
phases in the absorbing gas we allowed for the redshift of absorptions from different $J$ 
levels to be different.

\begin{figure*} 
\centering
\includegraphics[width=0.75\textwidth,angle=90]{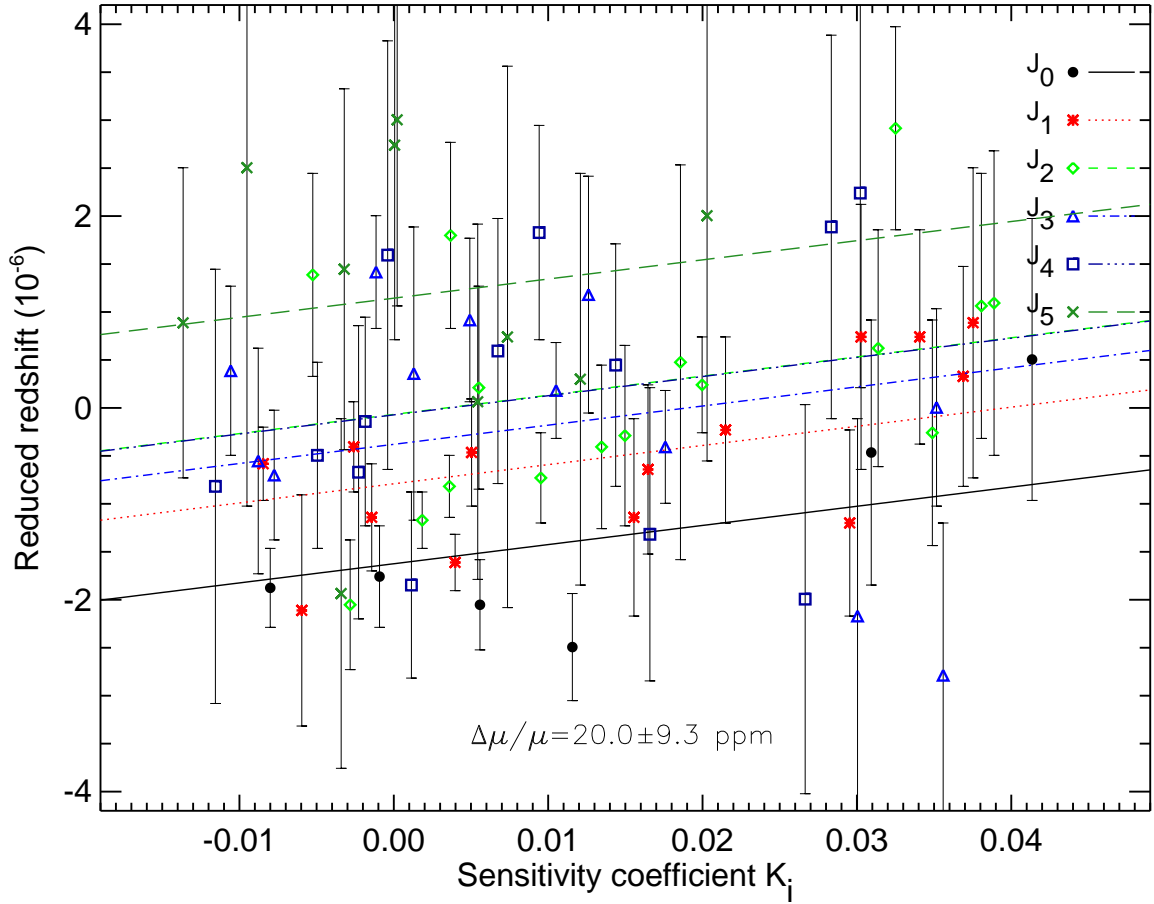}
\caption{Reduced redshift vs the  $\rm K_i$ for all the fitted $H_2$ lines.
Lines
from different $J$-levels are plotted with different symbols. The best fitted
linear line for different $J$-levels with the constraint that the slope should
be same is also shown. 
This analysis provides \dmm = 20.0+9.3 ppm, without any correction
for the velocity drift.
} 
\label{fig_zvsk_int}
\end{figure*}

In Fig.~\ref{fig_zvsk_int} we plot the reduced redshift vs $K$
for different transitions. The  different $J$-levels are
marked with different symbols and   the fitted line for  different $J$-levels are  shown   with different
line styles. The slope (i.e. \dmm) of
these lines is forced to be same.
The velocity drift as shown in Fig 1, was corrected for to give our final value: 
 \dmm\ = $+15.0\pm9.3$ ppm .

In the  second approach    we used both a single and  a two-component model.
The single component  model provides 
\dmm\ = $+15.6\pm6.9$ ppm . This  is    consistent with what we have
found above using $z$-vs-$K$ analysis.

The model with two components   accounts better  for the multi phase nature of the absorbing gas. In this case   $z$ and
$b$ of the two components  are forced to be  the same for different $J$-levels. 
The best-fit values is 
$ {\Delta}{\mu}/{\mu} = (-7.6 \pm  8.1_{\mathrm stat} \pm  6.3_{\mathrm sys}) $ ppm,
after correction for the velocity drift. 
The reduced $\chi^2$ is 1.177, that is slightly lower than the corresponding 
single component fit.  The two component model is marginally favored by the statistical indicators with respect to the single component and provides our favoured  value for \dmm. 

\begin{table*}
\caption{Selected values of \dmm\ from the literature\label{dmtab}}
\centering
\begin{tabular}{lllll}
\hline
\dmm\ & Ref. &  absorber & $z$ & QSO\\
$10^{-6}$\\
$4.3\pm7.2$                            & \citet{wm12}         &$\mathrm H_2$         &  3.025 &  Q0347$-$383\\ 
$0.3\pm3.2_{\rm stat}\pm1.9_{\rm sys}$ & \citet{King11}       & $\mathrm H_2$ and HD &  2.811 &  Q0528$-$250\\
$8.5\pm4.2$                            & \citet{vw11}         & $\mathrm H_2$ and HD &  2.059 &  J2123$-$005\\
$10.9\pm7.1$                           & \citet{King08}       &$\mathrm H_2$         &  2.595 &  Q0405$-$443\\
$3.7\pm14  $                           & \citet{Thompson09apj}&$\mathrm H_2$         &  2.595 &   Q0405$-$443\\
\hline
\end{tabular}
\end{table*}

Our measurement 
is consistent with a constant  $\mu$ over the last  $\approx$ 11  Gyr  within  one part in 10$^{5}$. 
This is consistent with  \dmm\ measurements in literature as reported in Table
\ref{dmtab}.
The measurement towards Q0528$-$250 
has the smallest estimated error,
there is however some concern on this measurement, due
to the fact that  \citet{King11} and \citet{Noterdaeme08}
derive molecular hydrogen column densities that differ
by a factor of 50. Further investigation of this system
is highly desirable. 
We note that three out of four UVES based measurements show positive values
of \dmm. 
However, since 
wavelength dependent drift, as observed by us,
could bias  \dmm\ measurements towards positive values so  this cannot  be taken 
as  evidence of variation until  the origin of the UVES velocity drifts is fully elucidated.

\balance

\section{Conclusions and future prospects}

The analysis of the first two lines of sight of the ESO Large Program dedicated to the study of the variability of the fundamental constants provided  results which are consistent with a null variation of the fine structure constant  $\alpha$ and of the proton-to-electron mass ratio $\mu$. Namely:
\\

\centerline{ 
 \daa=  $+1.3\pm   2.4_{\mathrm stat} \pm 1.0_{\mathrm sys}$ {\rm ppm} }

     and

\centerline{
\dmm\ = $-7.6 \pm  8.1_{\mathrm stat} \pm  6.3_{\mathrm sys} $ ppm }

\bigskip

 The analysis of the other  absorption systems towards the remaining lines of sight   is in progress. 
With the first analysis  we have confirmed the importance of accurate observational strategy targeted to minimize the systematics. In particular the use of the  solar spectrum obtained by regular asteroid observations  proved to be crucial  to check  the wavelength accuracy of the UVES spectrograph. This analysis revealed a systematic in the UVES wavelength scale with intra-order distortions  which  may have an  impact into   a possible signal in the variability of $\alpha$  and $\mu$.  A full characterization of these distortions is required in order to make a significant advance in the accuracy of these measurements.


\acknowledgements
P.B. acknowledges support from the Conseil Scientifique
de l'Observatoire de Paris.
S.A.L.'s work is supported by the grant DFG
Sonderforschungsbereich SFB 676 Teilprojekt C4.
P.M. and C.J.M. acknowledge the financial support of grant 
PTDC/FIS/111725/2009 from FCT (Portugal). C.J.M. is also supported by an FCT Research Professorship, contract reference IF/00064/2012.
M.T.M. thanks the Australian Research Council for funding under the Discovery Projects scheme (DP110100866).

\bibliographystyle{an}

\end{document}